# Characterisation of placental malaria in olive baboons (*Papio anubis*) infected with *Plasmodium knowlesi* H strain

**Barasa Mustafa[1,2]\* Gicheru Muita Michael[1], Kagasi Ambogo Esther[1,2],
Ozwara Suba Hastings[2]**

[1]Zoological Sciences Department, School of Pure and Applied Sciences, Kenyatta University, Nairobi
Department of Tropical and Infectious Diseases, Institute of Primate Research, Nairobi, Kenya

**ABSTRACT**

Pregnant women have increased susceptibility to malaria infection. In these women, malaria parasites are frequently found sequestered in the placental intervillous spaces, a condition referred to as placental malaria (PM). Placental malaria threatens the health of the mother and the child's life by causing still births and reduction in gestational age. An estimated 24 million pregnant women in Sub-Saharan Africa are at risk. Mechanisms responsible for increased susceptibility in pregnant women are not fully understood. Pregnancy malaria studies have been limited by the lack of a suitable animal model. This research aimed to develop a baboon (*Papio anubis*) model for studying PM. The pregnancies of three adult female baboons were synchronized and their gestational levels confirmed by ultrasonography. On the 150th day of gestation the pregnant baboons were infected with *Plasmodium knowlesi* H strain parasites together with four unilugravid control baboons. Parasitaemia was monitored from two days post inoculation until the 159th day of gestation when caesarean section was done on one baboon in order to obtain the placenta. Two baboons aborted their conceptus. Smears prepared from placental blood demonstrated the presence of *Plasmodium knowlesi* parasites in all the three sampled placentas. These new findings show that *P. knowlesi* sequesters in the baboon placenta. In addition, this study has characterized haemoglobin, eosinophil, Immunoglobulin G and Immunoglobulin M profiles in this model. Thus a non human primate (baboon) model for studying PM has been established. The established baboon – *P. knowlesi* model for studying human placental/pregnancy malaria now offers an opportunity for circumventing the obstacles experienced during human studies like having inadequate tissue for analysis, inaccurate estimation of gestational age, moral, ethical and financial limitations.

## INTRODUCTION

Many experiments have demonstrated that there is increased susceptibility to *P. falciparum* malaria infection in pregnant women (Menendez *et al.,* 2000; Malhotra *et al.,* 2002). In these women, malaria parasites frequently sequester in placental intervillous spaces, a condition referred to as placental malaria (PM). Malaria during pregnancy leads to many complications in women and their infants, threatening the lives of both the mother and the child. These complications include stillbirths, abortion, pre-term births, low birth weights, reduction in gestational age, anaemia and high fever (Menendez *et al.,* 2000; Steketee *et al.,* 2001). An estimated 24 million pregnant women in sub Saharan Africa are at risk of suffering from PM and prevalence may exceed 50% among primigravidae and secundigravidae in endemic areas (Phillips-Howard *et al.,* 1999). There is no vaccine and currently the best therapeutic measure involves the use of artemisinin-based combinational therapy.

Although PM is to an extent studied in pregnant women the studies have shortcomings resulting from confounding inherent variables such as the mother's health status, inaccurate estimation of gestational age, inadequate tissue for analysis, patient compliance problems, socio-economic conditions, moral, ethical and financial limitations (Moore *et al.,* 1999; Steketee *et al.,* 1996). As a result, many questions are not satisfactorily addressed during human studies. This study established and characterized a baboon model for studying placental malaria in baboons experimentally infected with *Plasmodium knowlesi* H strain parasites.

**\****Corresponding author:*
*Mustafa Barasa, MSc.*
*Zoological sciences Department,*
*School of Pure and Applied Sciences,*
*P. O. Box 43844, Kenyatta University, Nairobi, Kenya*
*Email: mustrech@yahoo.com*

Generated data demonstrated PM in the baboon model and will facilitate studies that will help in understanding the pathogenesis of PM in women and could help in designing of management and control strategies.

## MATERIALS AND METHODS

### Study design and experimental animals

Three adult female baboons were maintained in the baboon colony facility at the Institute of Primate Research (IPR) in the company of an adult male baboon for mating to occur. Ultrasound tests were used to confirm pregnancy status and gestation periods of the baboons. The baboons were infected together with four non pregnant control baboons with *Plasmodium knowlesi* blood stage, overnight cultured parasites (Ozwara *et al.*, 2003) on the gestation day 150. Each baboon received an inoculum of $1.0 \times 10^6$ parasites/ml in incomplete RPMI 1640. Following infection the baboons were finger pricked daily for parasitaemia determination (Ozwara *et al.*, 2003). At 5% level of parasitaemia baboons were intravenously injected with chloroquin sulphate at a dosage of 5 mg /kg daily for 3 days. Baboon groups used were as follows: (1) Pregnant infected; PAN 2724 (PAN means *Papio anubis*), PAN 2809, PAN 2859 and (2) Non pregnant infected; PAN 2870, PAN 3023, PAN 2911 and PAN 3035. The baboons were bled weekly for determination of haemoglobin and eosinophil levels. For all invasive procedures, the baboons were anaesthetised with ketamine hydrochloride (10mg/kg). PAN 2809 was taken for cesarean section seven days post infection while PAN 2724 and PAN 2859 aborted before cesarean sections could be performed.

### Cesarean section procedures and placental parasitaemia analysis

Caesarean section (CS) was performed (as described in IPR Standard Operating Procedures) in order to obtain intact sterile placental tissue from one baboon which can otherwise be consumed by the baboon. For PAN 2809 (pregnant infected baboon) the procedures were undertaken at 7 days post infection. Surgeons performed the abdominal incisions with minimum bleeding. Massaging of the baboons' abdomen was done and light pressure applied to the area at the top of the uterus. This manipulation was enough to expel the placenta. The freshly extracted placental tissue was placed in a sterile petridish with the maternal side facing up (Fig. 1). The infant delivered by CS was then euthanized (Fig. 2). Surgical blades were used to prick the maternal side of the placenta after wiping with cotton wool. Drops of blood were used to prepare thick and thin placental blood smears before staining and parasitaemia analysis as described before (Moore *et al.*,1999).

### Immunoglobulin G (IgG) and Immunoglobulin M (IgM) measurement

Immunoglobulin G (IgG) and Immunoglobulin M (IgM) were measured using ELISA in order to determine the level of humoral immune responses provoked in the baboons. Ninety six well flat bottomed ELISA microtiter plates were coated using 50 μl/well of carbonate bicarbonate buffer (pH 9.6) containing 108 (parasites used in antigen preparation) crude sonicated *P. knowlesi* parasite antigen. These were incubated overnight at 4°C for adsorption. Excess coating buffer was then flicked of and 100 μl/well of blocking buffer (3% BSA in PBS) added using a multichannel pipette followed by incubation for 1 hour at 37°C. The plates were then washed six times using wash buffer (0.05% tween in PBS) delivered by an automatic washing machine and sera samples added in triplicate (1:100 dilution in 50 μl/well) on ice before incubation for one hour as before. The plates were then washed as before and 50 μl/well of 1:2000 dilution of either antihuman Immunoglobulin G horseradish peroxidase or antihuman Immunoglobulin M horseradish peroxidase (HRP) added. One hour incubation at 37°C then followed after which the plates were washed ten times using washing buffer. Colour development was achieved by adding 50 μl/well of Tetramethylbenzidene (TMB) substrate and optical densities were read using a Dynatech MRX ELISA reader at 630 nm filter setting after 15 min of incubation at 37°C (Gicheru *et al.*, 1995).

### Statistical analysis

Mean values of parasitaemia, haemoglobin, eosinophil, IgG and IgM levels of pregnant infected group of baboons were compared with the values of control group of baboons (non pregnant infected baboons) using non parametric Mann-Whitney U analysis. Probability values of P<0.05 were considered significant.

## RESULTS

### Parasitaemia development, sequestration and induction of abortions

Development of peripheral parasitaemia was monitored on a daily basis in both animal groups. Pregnant baboons became parasitaemic (0.02%) beginning from day 2 post infection while non pregnant ones became parasitaemic (0.04%) from day 4 post infection. The mean peak peripheral parasitaemia in pregnant baboons (2.73%) and in non pregnant baboons (6.74%) differed significantly (Fig. 5 *[Supplementary data]*; P > 0.05).

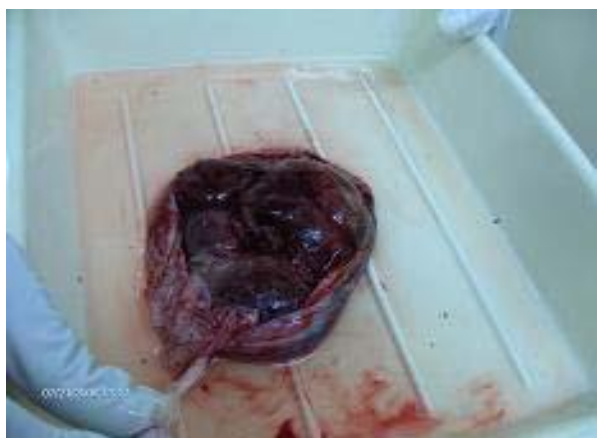

**Figure 1:** Placenta from PAN 2809. This placenta was pricked using a sterile blade and smears were made from the blood that oozed out.

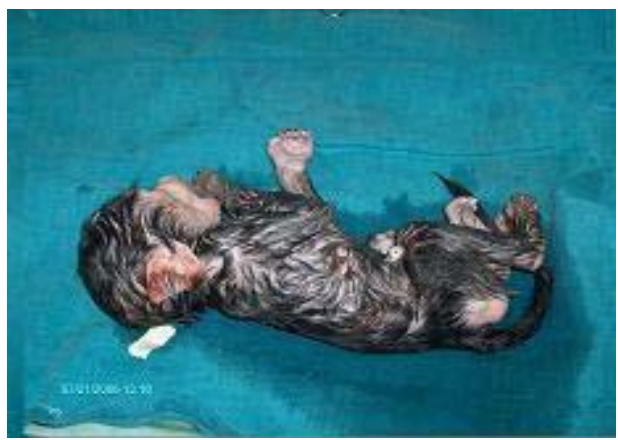

**Figure 2:** An infant delivered by cesarean surgery.

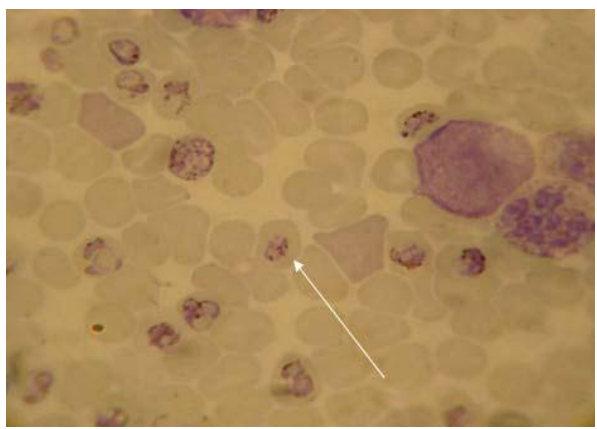

**Figure 3:** Placental blood smear from PAN 2859 representing a trophozoite stage of *P. knowlesi* as indicated by the arrow.

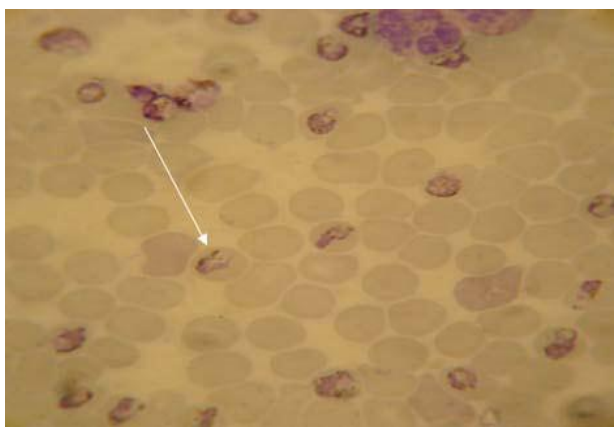

**Figure 4:** Placental blood smear from PAN 2724 representing a ring stage of *P. knowlesi* as indicated by the arrow.

Parasites were detectable in the placentas of all the three pregnant baboons in the experiment. Mean placental parasitaemia (11.9%) was 27 fold higher than the mean peripheral parasitaemia of 0.4% in pregnant baboons (Fig. 6 *[Supplementary data]*; P<0.05). The placentas of the baboons PAN 2859 and PAN 2724 were sampled following abortions by the baboons on days six and seven post infection respectively. Placental parasitaemia was 14.23% in PAN 2859, 18.30% in PAN 2724 and 0.74% in PAN 2809, sampled after cesarean section. Simultaneous peripheral parasitaemia in these baboons were 0.39%, 0.8% and 0.03% respectively. On the days when placental malaria was characterized, non pregnant baboons had over 6 fold higher peripheral parasitaemia than pregnant ones (Fig. 7 *[Supplementary data]*). When parasite stages were considered, rings (Fig. 4) and schizonts counted in the placenta (7.93% and 2.70% respectively) were found to be significantly (P<0.05) higher than in peripheral blood (0.13% and 0.19% respectively; P<0.05; Fig. 6). Trophozoite (Fig. 3) levels were not significantly different between placental (0.46%) and peripheral blood (0.10%; Fig. 6; P>0.05).

## Haemoglobin and Eosinophil levels

Pregnant baboons and non pregnant ones had comparable baseline and one week post infection levels of Hb concentration (baseline levels were 14.23 g/dl and 14.25 g/dl while week one post infection levels were 11.2 g/dl and 11 g/dl respectively). In pregnant baboons Hb levels ranged between 6 g/dl and 14.2 g/dl while in non pregnant baboons it ranged between 7.3 g/dl and 14.2 g/dl. Haemoglobin concentrations measured were significantly lower in pregnant baboons than in non pregnant baboons late during the infection (Fig. 8 *[Supplementary data]*; P<0.05). On average, pregnant baboons had higher baseline eosinophil levels than non pregnant baboons (177.33 compared to 112.35; Fig. 9 *[Supplementary data]*; P<0.05). Following infection, no eosinophils were observed in the pregnant group henceforth. The non pregnant baboons had a reduction by week 1 (from 112.25 to 22.75) followed by an increase by week 2 (from 22.75 to 80) only for eosinophils to disappear over week 3 and 4. In the pregnant group of baboons eosinophils ranged between 0 and 177.3 while in the non pregnant baboons they ranged between 0 and 112.25.

### Immunoglobulin G and Immunoglobulin M levels

All baboons experienced gradual increases in Ig G levels over the first two weeks of infection. In pregnant baboons levels ranged between 1.08 and 1.47 while in non pregnant baboons levels ranged between 1.45 and 3.36. There was a significant difference in IgG levels between pregnant and non pregnant baboons over the entire period of experimentation with non pregnant baboons producing higher IgG responses than pregnant baboons (Fig. 10 *[Supplementary data]*; P<0.05). Generally all baboons had rapid increases in malaria antigen specific Ig M titers by one week post inoculation. In pregnant baboons IgM levels ranged between 0.26 and 1.08 while in non pregnant baboons levels ranged between 0.27 and 1.05. Comparison of means of pregnant and non pregnant baboons revealed that differences in IgM production were not significant (Fig. 11 *[Supplementary data]*; P > 0.05).

### DISCUSSION

The aim of this work was to develop a baboon (nonhuman primate) model for studying human pregnancy malaria. Placental parasitaemia on average was over 20 fold higher in the placenta than peripheral parasitaemia. This was clear evidence of sequestration of *P. knowlesi* in the baboon placentas. On the days when PM was characterized, non pregnant control baboons had 6 fold higher peripheral parastaemia than pregnant ones. The difference in peripheral parasitaemia between the two groups is likely to be as a result of sequestration in the placenta of pregnant baboons. The sequestration of *Plasmodium* parasites in placentas has been demonstrated in the murine (Jayekumar and Moore, 2006; Vinayak *et al.,* 1986; Desowitz *et al.,* 1989) and rhesus monkey (Das Gupta, 1934) models before. The presence and sequestration of malaria parasites in the placental tissues, accompanied by low peripheral parasitaemia is one definitive feature of placental malaria in humans (Brabin, 1983). In a typical case of placental malaria, at times there is placental parasitisation even when simultaneous peripheral parasitaemia is negative (Menendez *et al.,* 1995).

Pregnant baboons also seemed to be more highly susceptible to infection than non pregnant ones since they became parasitaemic two days earlier than non pregnant baboons. Pregnant women are known to have increased susceptibility to malaria infection and are more likely to develop symptomatic malaria than their non pregnant counterparts (Davison *et al.,* 1998). This sequestration finding could suggest that the baboon placenta may contain molecules that mediate parasite adherence to the placental tissue like chondroitin sulphate A which have been reported in humans (Fried *et al.,* 1996; Sartelet *et al.,* 2000).

Placental blood smears contained ring stages as the predominant parasite developmental stage. There is therefore a strong possibility that stages that adhered to the placental tissue were mainly trophozoites and schizonts. These schizonts and trophozoites could have preferentially adhered to placental tissue hence failure of their detection in blood smears prepared from the placenta. *Plasmodium falciparum* can be distinguished from other *Plasmodium* species that infect humans because only immature ring-infected erythrocytic forms circulate in the peripheral blood (Miller *et al.,* 1998).

Two of three pregnant baboons in this study experienced abortion at about one week post infection. The baboon that did not abort had over 40-fold lower placental parasitaemia compared to the others. Although more animals (a greater sample size) are required to confirm that *P. knowlesi* infection in pregnant baboons causes abortion, this outcome could suggest that high placental parasitaemia in *P. knowlesi* infected baboons results into abortion. This study also shows that the heavier the placental parasitisation, the greater the chances of abortion since the baboon with the lowest placental parasitaemia was the only one that did not abort thus allowing cesarean delivery. Malaria induced abortions have previously been demonstrated in the murine (Jayekumar and Moore, 2006) and rhesus monkey models of malaria (Davison *et al.,* 1998). Malaria during pregnancy leads to many complications in women and their infants like anaemia, hypoglycemia, pulmonary edema, congenital infection and increased foetal and maternal mortality and morbidity, renal failure, intrauterine growth retardation (IUGR), cerebral malaria, abortion and preterm delivery (Menendez *et al.,* 1995). In areas of unstable prevalence of malaria, pregnant women are highly susceptible to *P. falciparum* malaria infection and have 2-20 fold higher mortality rates than the non pregnant ones.

Pregnant baboons had reduced levels of haemoglobin in the later stages of infection. Since their lower haemoglobin concentration could suggest low RBC density, this could mean that pregnant baboons had slower or poor recovery from infection-mediated RBC reduction. Reduction in haemoglobin levels is a well characterized occurance in malaria infection (Davison *et al.,* 1998). In his study Davison *et al.,* (1998) with the *P. coatneyi*-rhesus monkey model of malaria pregnant monkeys became anaemic following infection in the first trimester. Reduction in haemoglobin was also detected in the *P. berghei*-white rat model and also in *P. chabaudi*-B6 mice model (Jayekumar and Moore, 2006). In the latter study, pregnant mice had accelerated anaemia than non pregnant ones. Pregnant baboons had higher baseline eosinophil levels than non pregnant baboons. Following infection, very low eosinophils were observed in the pregnant group henceforth yet they were still detected in non pregnant baboons.

Eosinophils have large eosinophilic granules and non segmented or bilobed nucleus. They are few in tissues except in certain types of inflammation and allergies. Eosinophils participate in inflammatory reactions and immunity to some parasites. Baboon eosinophils could have infiltrated the placental tissues leading to a drastic reduction (of eosinophils) in the peripheral circulation. This is reported in human placental malaria (Matteelli *et al.,* 1997). Eosinophils could infiltrate the placenta as a result of parasite induced inflammatory reaction.

In this study, the non pregnant group of baboons raised significantly higher titers of IgG than the pregnant group of baboons. This result could mean that, through some mechanism, peripheral production of IgG is suppressed in pregnant *P. knowlesi* infected baboons. Generally comparisons done between the pregnant and control non pregnant groups of baboons showed that no significant differences existed between the two groups in terms of IgM synthesis. This means that IgM mediated immunity could be equally effective in both groups. Pregnancy causes a number of physiological changes that affect the way the *Plasmodium* parasite invades its host. Down regulation of normal maternal immune response is necessary to prevent rejection of the conceptus. Cell mediated immunity (Th1) is particularly suppressed during pregnancy, and the mother is increasingly reliant on humoral immunity (Th2) for protection (Samak, 2004). This could be the reason why pregnant baboons had IgM reponses that were comparable to non pregnant baboons.

This study has established a baboon (non human primate) model for studying placental malaria. The study has demonstrated that *P. knowlesi* sequesters in the baboon placenta and sequestration primarily occurs in mature parasite stages. The study has also shown that placental malaria in the baboon model is associated with abortion and reduction in peripheral haemoglobin and eosinophils. The developed model will facilitate studies that will help in understanding the pathogenesis of PM in women and could help in designing of management and control strategies.

## ACKNOWLEDGEMENT

We are grateful to the Animal Resources Department at the Institute of Primate Research for providing the baboons and other support during the study. This study was supported by the research capability strengthening WHO (Grant Number: A 50075) for malaria research in Africa under the Multilateral Initiative on Malaria /Special Programme for Research and Training in Tropical Diseases (WHO-MIM/TDR).

## REFERENCES

Brabin BJ (1983) An analysis of malaria infection in Africa. *Bull World Health Org.*, **61**:1005-1016.

Das Gupta BM (1934) Malarial infection in the placenta and transmission to the foetus. *Indian Med. Gaz.*, **74**: 397-399.

Davison BB, Cogswell FB, *et al.* (1998) *Plasmodium coatneyi* in the rhesus monkey (*Macaca mulatta*) as a model of malaria in pregnancy. *Am. J. Trop. Med. Hyg.*, **59**: 189-201.

Desowitz R, Shida K, *et al.* (1989) Characterisation of a model of malaria in the pregnant host: *Plasmodium berghei* in the white rat. *Am. J. Trop. Med. Hyg.*, **41**: 630-634.

Fried M and Duffy PE (1996) Adherence of *Plasmodium falciparum* to chondroitin sulfate A in the human placenta. *Science*, **272**: 1502-1504.

Jayakumar P and Moore JM (2006) Murine malaria infection induces fetal loss associated with accumulation of *Plasmodium chabaudi* AS infected erythrocytes in the placenta. *Infect. Immun.*, **74**: 2839-2848.

Malhotra M, Sharma JB, *et al.* (2002). Maternal and perinatal outcomes in varying degrees of anaemia. *Int. J. Gynaecol. Obstet.*, **79**: 93-100.

Matteelli A, Caligaris S, *et al.* (1997) The placenta and malaria. *Ann. Trop. Med. Parasitol.*, **91**: 803-810.

Menendez C (1995) Malaria during pregnancy: a priority area of malaria research and control. *Parasitol. Today*, **11**:178-183.

Menendez C, Ordi J, *et al.* (2000) The impact of placental malaria on gestational age and birth weight. *J. Infect. Dis.*, **181**: 1740-1745.

Miller LH and Smith JD (1998). Motherhood and malaria. *Nat. Med.*, **4**: 1244-1245.

Moore JM, Nahlen BL, *et al.* (1999) Immunity to placental malaria. I. Elevated production of interferon-gamma by placental blood mononuclear cells is associated with protection in an area with high transmission of malaria. *J. Infect. Dis.*, **179**: 1218-1225.

Ozwara HS, Langermans JA, *et al.* (2003) Experimental infection of the olive baboon (*Papio anubis*) with *Plasmodium knowlesi*: severe disease accompanied by cerebral involvement. *Am. J. Trop. Med. Hyg.*, **69**: 188-194.

Phillips-Howard PA (1999) Epidemiological and control issues related to malaria in pregnancy. *Ann. Trop. Med. Parasitol.*, **1**: 11-17.
Samak AC (2004) Pregnancy in malaria: an overview. *McGill J. Medicine*, **8**: 66-71.

Sartelet H, Garraud O, *et al.* (2000) Hyperexpression of ICAM-1 and CD 36 in placentas infected with *Plasmodium falciparum*: a possible role of these molecules in sequestration of infected red blood cells in placentas. *Histopathology*, **36**: 62-68.

Steketee RW, Nahlem BM, *et al.* (2001) The burden of malaria in pregnancy in malaria-endemic areas. *Am. J. Trop. Med. Hyg.*, **64**: 28-35.

Steketee RW, Wirima JJ, *et al.* (1996). The problem of malaria and malaria control in pregnancy in sub-Saharan Africa. *Am. J. Trop. Med. Hyg.*, **55**: 2-7.

Vinayak V, Pathak G, *et al.* (1986). Influence of malarial infection on the maternal foetal relationship in pregnant mice. *Aust. J. Exp. Bio. and Med. Sci.*, **64**: 223-227.